# Optimization of Energy Resolution and Pulse Shape Discrimination for a CLYC Detector with Integrated Digitizers

**Tao Xue**[a,b] **, Jinfu Zhu**[a,b]**, Jingjun Wen**[a,b]**, Jirong Cang**[a,b,c*]**, Zhi Zeng**[a,b]**, Liangjun Wei**[d]**, Lin Jiang**[a,b]**, Yinong Liu**[a,b] **and Jianmin Li**[a,b]

[a] *Key Laboratory of Particle & Radiation Imaging, Ministry of Education,100084, Beijing, China*
[b] *Department of Engineering Physics, Tsinghua University,100084, Beijing, China*
[c] *Department of Astronomy, Tsinghua University,100084, Beijing, China*
[d] *NUCTECH Company Limited,100084, Beijing, China*

*E-mail*: cangjirong@mail.tsinghua.edu.cn

ABSTRACT: Sufficient current pulse information of nuclear radiation detectors can be retained by direct waveform digitization owing to the improvement of digitizer's performance. In many circumstances, reasonable cost and power consumption are on demand while the energy resolution and PSD performance should be ensured simultaneously for detectors. This paper will quantitatively analyse the influence of vertical resolution and sampling rate of digitizers on the energy resolution and PSD performance. The energy resolution and PSD performance can be generally optimized by improving the sampling rate and ENOB (effective number of bits) of digitizers. Several integrated digitizers, with sampling rates varying from 100 MSPS to 500 MSPS and vertical resolution ranging from 12-Bit to 16-Bit, were designed and integrated with a CLYC detector for verifications. Experimental results show good accordance with theoretical calculations. The conclusion can give guidance to designs of digitizes for similar applications in need of optimizing the energy resolution and PSD performance, and help to choose proper digitizers for different requirements.

KEYWORDS: Energy Resolution; Pulse Shape Discrimination; Integrated digitizer; CLYC scintillator.

---

* Corresponding author.

**Contents**

## 1. Introduction

$Cs_2LiYCl_6$ (CLYC) scintillator has the ability to discriminate the fast neutron, thermal neutron, and gamma-ray. It is a potential counterpart for $^3$He proportional counter tube [1] [2] and can be used as high resolution gamma-ray and neutron spectrometers due to its exceptional PSD (Pulse Shape Discrimination) performance [3]. Digital PSD algorithms [4] have been widely used in radiation detectors such as $LaBr_3$ scintillator [5], plastic scintillators [6], and organic scintillators [7] [8]. PSD FOM (Figure Of Merit) is generally used to evaluate the performance of the particle discrimination, e.g. GRR (Gamma Rejection Ratio) is defined as $0.5 \times erfc\left(2\sqrt{\ln 2} \times FOM\right)$ [9].

With the rapid development of ADC (Analog-to-Digital Converter) and FPGA (Field Programmable Gate Array), the vertical resolution and sampling rate of an integrated digitizer have been substantially increased [10] [11] [12]. Generally, digitizers with higher performance will result in more cost and power consumption. It is critical to understand the correlations between the resolution or/and sampling rate of a digitizer and the energy resolution or/and PSD FOM of a radiation detector. Such correlations will be analysed and verified with different digitizers. The conclusion is useful to choose or design digitizers in specific applications.

This paper will demonstrate methods of quantitative analysis on energy resolution and PSD FOM in Section II. Section III will introduce the CLYC detection system and setup of experiments. The results will be illustrated in Section IV and the conclusion and future work will be discussed in Section V.

## 2. Method

### 2.1 The quantitative analysis of energy resolution

For a typical $^{137}$Cs gamma energy spectrum, as depicted in Figure 1, the energy resolution is defined as equation (1). As indicated in the Gaussian fit of the full-energy peak, $E$ is the peak location, and $\Delta E$ is FWHM (Full Width at Half Maximum) of deposited energy distribution, $\sigma_E^2$ is the variance of deposited enengy, and $\eta$ is the energy resolution.

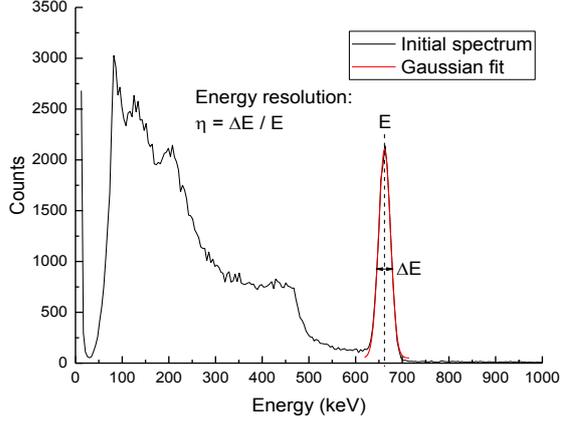 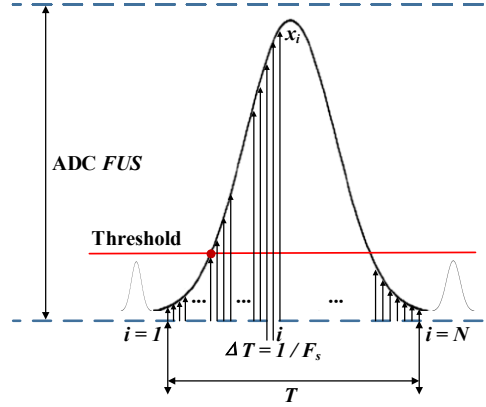

**Figure 1.** The typical $^{137}$Cs gamma energy spectrum.   **Figure 2.** The current pulse integration.

For a current pulse from detector, its integrated charge $Q$ is proportional to the deposited energy $E$ of an incident particle. The uncertainties of the charge are generally contributed by intrinsic statistical fluctuation (scintillator) and electronic noise (Photomultiplier Tube, preamplifier and digitizer). When the signal spans more than a few LSBs (the Least Significant Bits) or the input-referred noise is larger than one-half LSB of ADC [13], the quantized noise from the digitizer can be considered uncorrelated with preamplifiers and detectors. The total uncertainty of the charge consists of the contribution from the digitizer and other parts, such as statistical fluctuation and preamplifiers, etc., as described in equation (2).

$$\eta = \frac{\Delta E}{E} \approx \frac{2.355\sigma_E}{E} = \frac{2.355\sigma_Q}{Q} \quad (1)$$

$$\sigma_Q^2 = \sigma_{digitizer}^2 + \sigma_{other}^2 \quad (2)$$

The current pulse integration is illustrated in figure 2, which is directly sampled by an ADC followed the current amplifier. $x_i\,(unit:V)$ is the amplitude for $i\,th$ sampling point, $\Delta T\,(unit:s)$ is sampling period and $F_S$ is sampling rate ($\Delta T = 1/F_S$). As shown in equation (3), the integrated charge $Q\,(unit:C)$ is obtained by the summation of $N$ sampling points ($N = T/\Delta T$), where $A\,(unit:ohm)$ is the I-V (current to voltage) gain of preamplifier before ADC.

$$Q = \sum_{i=1}^{N} \Delta T \times \frac{x_i}{A} = \sum_{i=1}^{N} \frac{1}{F_S} \times \frac{x_i}{A} \quad (3)$$

Refer to the error propagation formula, for $u = u(x_1, x_2, x_3, ..., x_N)$, if $x_1, x_2, x_3, ..., x_N$ is uncorrelated, the total variance is $\sigma_u^2 = \sum_{i=1}^{N} \left(\frac{\partial u}{\partial x_i}\right)^2 \sigma_{x_i}^2$. According to the ADC quantized error formula [14], $\sigma_{x_i}^2$ contributed by the digitizer is estimated to $\left(\frac{FUS}{2^{ENOB}}\right)^2 \frac{1}{12}$. The variance of charge contributed by the digitizer can be described as equation (4), where *FUS* is the full-scale differential input of ADC, ENOB is the effective number of bits of ADC and *T* is the integration time.

$$\sigma_{digitizer}^2 = \left(\frac{\Delta T}{A}\right)^2 \sum_{i=1}^{N} \sigma_{x_i}^2 = \frac{1}{F_s} \left(\frac{FUS}{A \times 2^{ENOB}}\right)^2 \frac{\Delta T \times N}{12} = \frac{1}{F_s} \left(\frac{FUS}{A \times 2^{ENOB}}\right)^2 \frac{T}{12} \quad (4)$$

The energy resolution contributed by the digitizer is shown in equation (5).

$$\eta \approx \frac{2.355 \sigma_Q}{Q} = \frac{2.355 \left(\sqrt{\sigma_{other}^2 + \frac{1}{F_s} \left(\frac{FUS}{A \times 2^{ENOB}}\right)^2 \frac{T}{12}}\right)}{Q} \quad (5)$$

**2.2 The quantitative analysis of pulse shape discrimination**

The typical pulses of gamma-ray and neutron are shown in figure 3. The PSD ratio *R* is calculated as equation (6), where $Q_L$ and $Q_S$ are long delay and short prompt integrations of charge respectively. The uncertainties of $Q_L$ and $Q_S$ caused by digitizers are uncorrelated, substituting equation (4), the variance of *R* contributed by the digitizer can be calculated using equation (7), where $T_L$ and $T_S$ are delay and prompt integration time respectively.

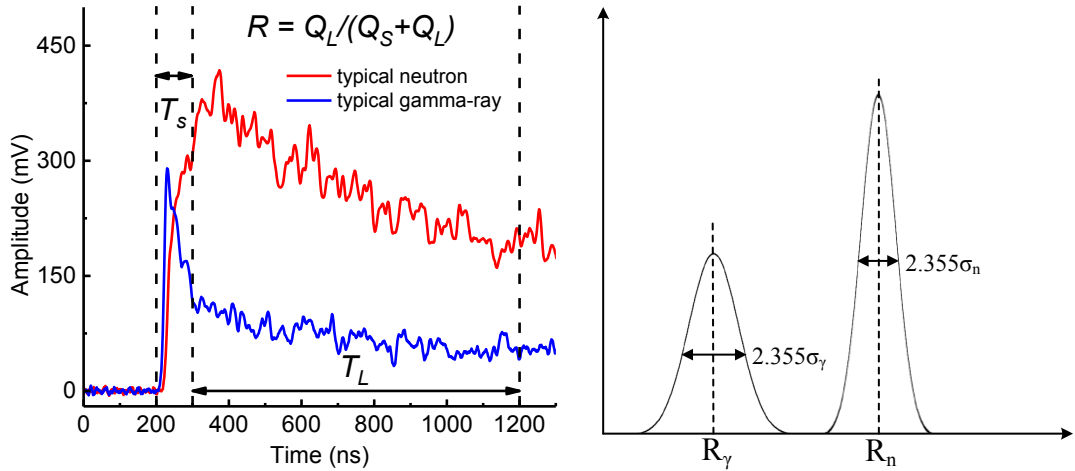

**Figure 3.** The typical pulses of neutron and gamma-ray.   **Figure 4.** The PSD ratio distribution.

$$R = \frac{Q_L}{Q_L + Q_S} \quad (6)$$

$$\sigma^2_{R(digitizer)} = \left(\frac{\partial R}{\partial Q_L}\right)^2 \sigma^2_{Q_L} + \left(\frac{\partial R}{\partial Q_S}\right)^2 \sigma^2_{Q_S} = \frac{Q_S^2 \sigma^2_{Q_L} + Q_L^2 \sigma^2_{Q_S}}{(Q_L + Q_S)^4} = \frac{T_L Q_S^2 + T_S Q_L^2}{12 \times (Q_L + Q_S)^4 F_s} \left(\frac{FUS}{A \times 2^{ENOB}}\right)^2 \quad (7)$$

Figure 4 demonstrates a typical distribution of PSD ratio for neutron and gamma-ray in CLYC detectors, where $R_n$ and $R_\gamma$ is the mean value of neutron and gamma-ray respectively. The PSD FOM is defined as equation (8).

$$FOM = \frac{|R_\gamma - R_n|}{2.355(\sigma_{R_\gamma} + \sigma_{R_n})} = \frac{|R_\gamma - R_n|}{2.355\left(\sqrt{\sigma^2_{R_\gamma(digitizer)} + \sigma^2_{R_\gamma(other)}} + \sqrt{\sigma^2_{R_n(digitizer)} + \sigma^2_{R_n(other)}}\right)}$$

$$= \frac{|R_\gamma - R_n|}{2.355\left(\sqrt{\left[\frac{T_L Q_S^2 + T_S Q_L^2}{12 \times (Q_L + Q_S)^4 F_s}\left(\frac{FUS}{A \times 2^{ENOB}}\right)^2\right]_{R_\gamma(digitizer)} + \sigma^2_{R_\gamma(other)}} + \sqrt{\left[\frac{T_L Q_S^2 + T_S Q_L^2}{12 \times (Q_L + Q_S)^4 F_s}\left(\frac{FUS}{A \times 2^{ENOB}}\right)^2\right]_{R_n(digitizer)} + \sigma^2_{R_n(other)}}\right)}$$
(8)

## 3. Experiments

The detection system consists of a 25.4 mm diameter and 25.4 mm height CLYC scintillator with 95% enrichment of 6Li from SCIONIX, a R6231-100 PMT from Hamamatsu and readout electronics. As sketched in figure 5, PMT readout electronics is stacked with the high voltage power board, preamplifier (~50 MHz bandwidth), ADC board, ZYNQ board, POE (Power Of Ethernet) board, and User Interface board [15]. The PMT is connected with the readout electronics via a dedicated socket. The electronics system is shielded and supported by an aluminum cylinder.

Seven integrated digitizers have been designed for verifications. As summarized in table 1, the ENOBs from the datasheets are listed. Moreover, the ENOB at the reference frequency from the datasheet were also measured using a standard sinusoidal signal according to IEEE 1241-2000 standard. The degradation of the measured ENOB, compared with the manual reference, for integrated digitizers is most likely caused by the jitter of ADC's sampling clock. The full-scale voltage and the typical power consumptions of ADCs are also listed. Generally, the higher the ADC sampling rate, the greater the power consumption.

Different gamma-ray sources ($^{57}$Co, $^{137}$Cs and $^{60}$Co) are used for energy calibration and resolution calculations, a moderated neutron source with $^{252}$Cf and polyethylene is deployed to evaluate the PSD performance.

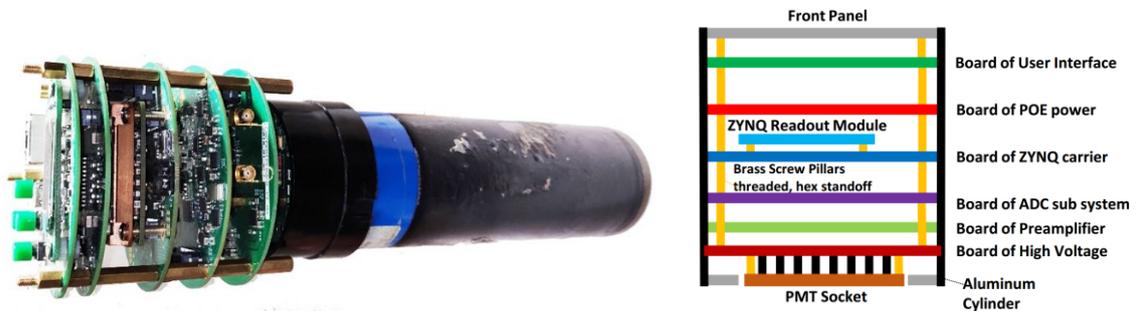

**Figure 5.** The integrated PMT readout electronics

Table 1. Parameters of integrated digitizers

| Digitizer | Speed (MSPS) / Resolution (Bit) | ENOB (datasheet) (Bit) | ENOB measured (Bit) | ADC full-scale voltage (V) | Power dissipation (mW / Channel) |
|---|---|---|---|---|---|
| ISLA212P50 | 500 / 12 | 11.27 @ 30 MHz | 10.70 | 1.8 | 858 |
| AD9642 | 250 / 14 | 11.50 @ 30 MHz | 10.30 | 1.75 | 360 |
| AD9634 | 250 / 12 | 11.20 @ 30 MHz | 10.00 | 1.8 | 360 |
| AD9265 | 125 / 16 | 12.80 @ 2.4 MHz | 10.70 | 1.8 | 439 |
| AD9255 | 125 / 14 | 12.70 @ 2.4 MHz | 11.00 | 1.8 | 437 |
| AD9233 | 125 / 12 | 11.40 @ 2.4 MHz | 10.38 | 1.8 | 415 |
| AD9253 | 100 / 14 | 12.10 @ 9.7 MHz | 11.25 | 1.8 | 101 |

## 4. Results

### 4.1 Energy resolution

The energy resolutions of the system at 662 keV measured with various digitizers are delineated as the circle points in figure 6. According to equation (5), combined with sampling rate and ENOB (effective number of bits) of digitizers, $F_S \times 4^{ENOB}$ is defined to represent the performance of digitizers. Basically, the sampling rate of digitizer should be larger than the Nyquist bandwidth of the input signal. The increase in the sampling rate as a factor of four can be achieved at the expense of decreasing one ENOB. As expected, the energy resolution becomes better with the increase of the sampling rate and vertical resolution. The solid line is the fitting result of how the energy resolution varies with $F_S \times 4^{ENOB}$, of which the fitting parameters are also shown in figure 6.

Intrinsic energy resolution without digitizer's contribution is estimated to be ~4.56 % at 662 keV according to the fitting results. When $F_S \times 4^{ENOB}$ is more than $3 \times 10^8$ MHz (e.g. 300 MSPS 10-Bit ENOB), the energy resolution will be approximated to the intrinsic energy resolution, and further enhancement of digitizer benefits a limited part for energy resolution. For some circumstance, such as portable gamma-ray spectrometer and radiation pagers, where the demand of energy resolution is ~5%, a low-cost digitizer (e.g. 100 MSPS 9-Bit ENOB) is sufficient, system cost and power consumption (not only the ADC, but also the associated electronics, such as clock generator, driver, buffer and FPGA) will be dramatically saved.

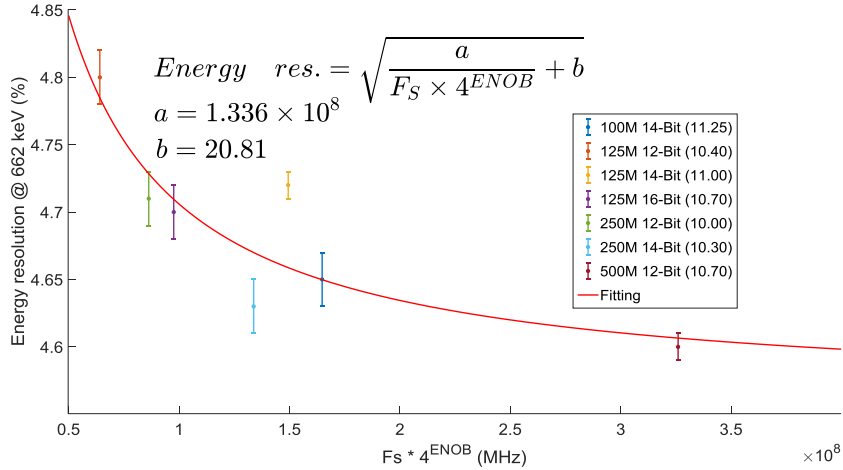

**Figure 6.** The energy resolution varies as $F \times 4^{ENOB}$.

## 4.2 PSD FOM

Three typical PSD ratio distributions and FOM calculations are portrayed in figure 7 with a moderated neutron source with $^{252}$Cf and polyethylene. They are measured by 500 MSPS, 250 MSPS and 125 MSPS digitizers with theoretical 12-Bit vertical resolution. As observed, the FOM becomes better with the increase of the sampling rate. The region pointed by thermal neutrons is from $^{6}$Li(n,α)t (Q = +4.79 MeV), which has a Gamma Equivalent Energy (GEE) of 3.2 MeV (67% conversion efficiency) approximately. In order to cover the region from thermal neutrons, the hits with GRR ranged from 1 to 4 MeV are selected to calculate the PSD FOM.

Similar in Sec 4.1, $F_S \times 4^{ENOB}$ is defined to represent the performance of digitizers. The correlation of PSD FOM and ADC performance is shown in figure 8. Equation (8) was used to fit the measured results and the results are shown in figure 8. The FOM improves slightly when $F_S \times 4^{ENOB}$ is more than $3 \times 10^8$ MHz (e.g. 300 MSPS 10-Bit ENOB). If GRR need to be less than 10$^{-12}$, FOM should be larger than 3.0 and the $F_S \times 4^{ENOB}$ should be greater than $2.98 \times 10^7$ MHz (e.g. 100 MSPS 9.1-Bit ENOB). Based on the fitted results, the intrinsic FOM is estimated to be ~4.1 assuming an ideal performance of ADC, which means an intrinsic GRR ~ $2.346 \times 10^{-22}$. Similarly, appropriate digitizer with moderate cost and power consumption can be selected based on our method for dedicated requirement of FOM or GRR.

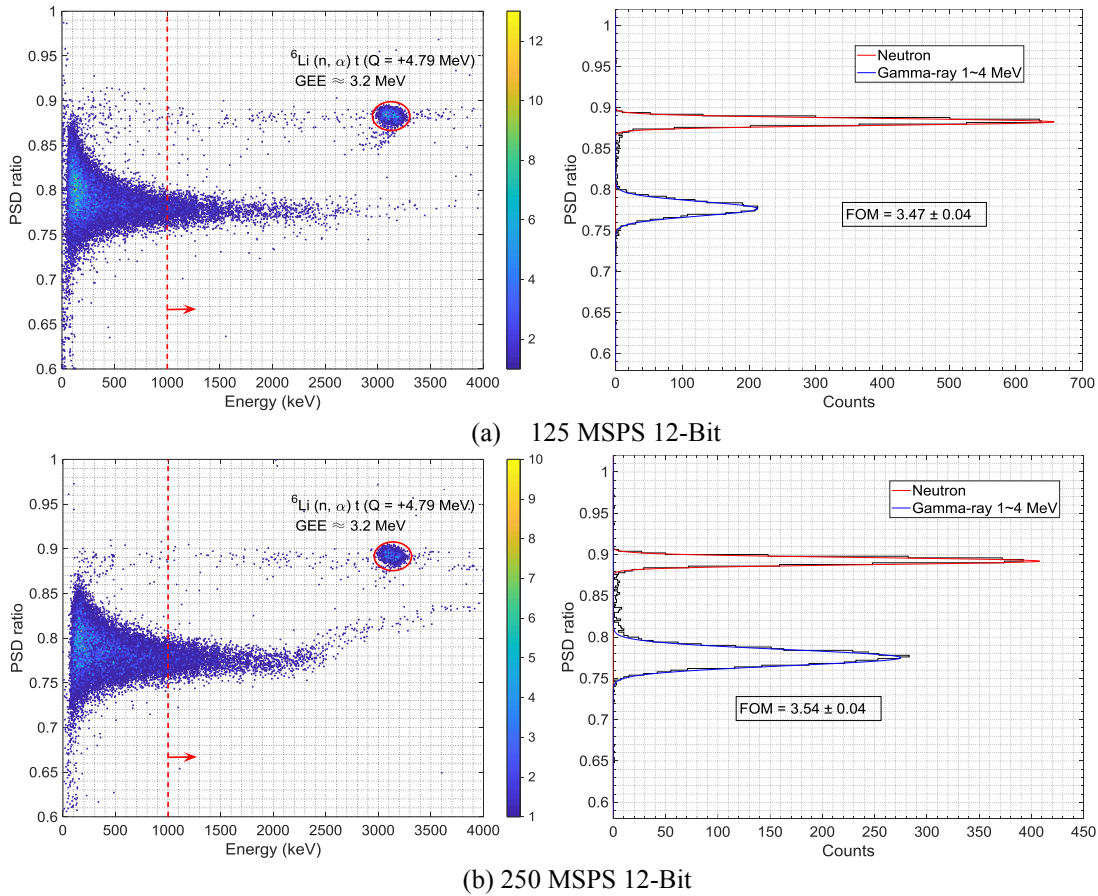

(a) 125 MSPS 12-Bit

(b) 250 MSPS 12-Bit

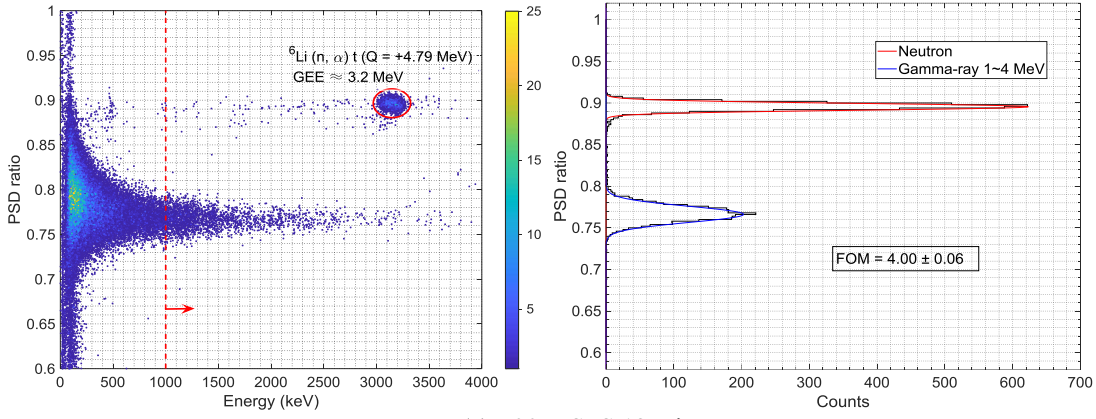

(c) 500 MSPS 12-Bit

**Figure 7.** The PSD ratio distribution of hits with 1 ~ 4 MeV GRR.

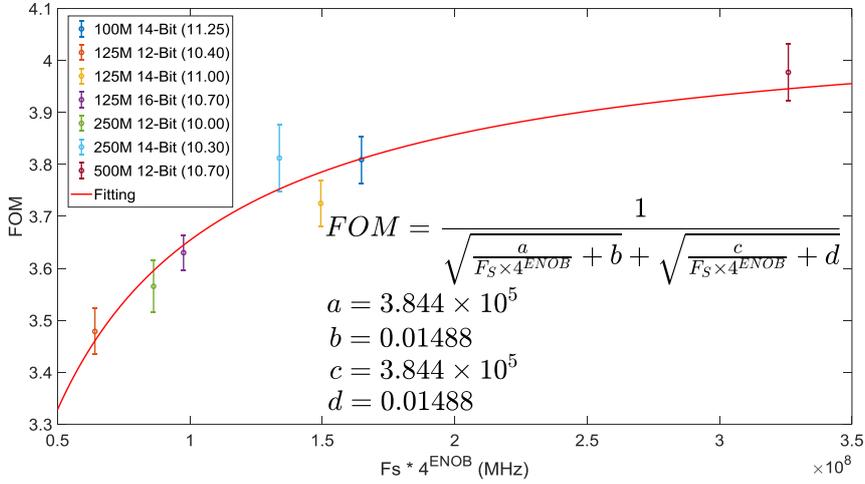

**Figure 8.** The FOM varies as $F_S \times 4^{ENOB}$.

## 5. Conclusions and future work

In this paper, the influences of the vertical resolution and sampling rate of digitizer on the energy resolution and PSD performance are quantitatively calculated. Seven digitizers with sampling rate ranging from 100 MSPS to 500 MSPS and vertical resolution varying from 12-Bit to 16-Bit are designed and integrated with CLYC detector for the verification of our calculation method. Experimental results are fitted and in good accordance with the calculation results. This paper contributes an effective guidance to choose proper digitizers to compromise with cost, power consumption, and practical performance, such as portable neutron and gamma-ray spectrometer, radiation pagers, high-energy particle detection in outer space, and thousands of PMTs used in discrimination of Cherenkov and scintillation light for neutrinos detection, etc.

In the future, in order to improve the energy resolution or PSD performance, intrinsic fluctuation from detectors and electronic noise from preamplifiers or PMTs will be intensive studied.

**Acknowledgments**

This work is supported by the National Key Research and Development Program of China (2017YFA0402202).

We would like to thank those who collaborated on the CDEX, and also thank Professor Yulan Li, Qian Yue, Litao Yang, Hao Ma and Zhi Deng for their support, information sharing and plenty discussions over the years at the Tsinghua University DEP (Department of Engineering Physics).

We are grateful for the patient help of Yu Xue, Wenping Xue, and Jianfeng Zhang. They are seasoned full-stack hardware technologist with wealth experience of solder and rework in the electronics workshop at DEP.

**References**

[1] J. Glodo, R. Hawrami, and K.S. Shah, *Development of $Cs_2LiYCl_6$ scintillator*, *Journal of Crystal Growth.* **379** (2013) 3.

[2] E. V. D. van Loef et al., *Optical and Scintillation Properties of $Cs_2LiYCl_6 : Ce^{3+}$ and $Cs_2LiYCl_6 : Pr^{3+}$ Crystals*, *IEEE Trans. Nucl. Sci.* **52** (2005) 5.

[3] William M. Higgins et al., *Bridgman growth of $Cs_2LiYCl_6$:Ce and $^6$Li-enriched $Cs_2LiYCl_6$:Ce crystals for high resolution gamma ray and neutron spectrometers*, *Journal of Crystal Growth* **312** (2010) 8.

[4] M. Nakhostin, *A General-Purpose Digital Pulse Shape Discrimination Algorithm*, *IEEE Trans. Nucl. Sci.* **66** (2019) 5.

[5] J. Cang et al., *Optimal design of waveform digitisers for both energy resolution and pulse shape discrimination*, *Nucl. Instrum Meth.* **A 888** (2018) 96 [arXiv: 1712.05207].

[6] E. V. Loef et al., *Plastic Scintillators With Neutron/Gamma Pulse Shape Discrimination*, *IEEE Trans. Nucl. Sci.* **61** (2014) 1.

[7] A. Favalli et al., *Pulse Shape Discrimination Properties of Neutron-Sensitive Organic Scintillators*, *IEEE Trans. Nucl. Sci.* **60** (2013) 2.

[8] M. Flaska et al., *Influence of sampling properties of fast-waveform digitizers on neutron-gamma-ray, pulse-shape discrimination for organic scintillation detectors*, *Nucl. Instrum Meth.* **A 729** (2013) 11.

[9] B. S. McDonald et al., *A wearable sensor based on CLYC scintillators*, *Nucl. Instrum Meth.* **A 821** (2016) 11.

[10] T. Xue et al., *The Design and Data-Throughput Performance of Readout Module Based on ZYNQ SoC*, *IEEE Trans. Nucl. Sci.* **65** (2018) 5.

[11] T. Xue et al., *The prototype design of integrated base for PMT with pulse digitalization and readout electronics*, 2018 *JINST* **13 T06007**.

[12] J. Zhu et al., *Preliminary Design of Integrated Digitizer Base for Photomultiplier Tube*, *IEEE Trans. Nucl. Sci.* **66** (2019) 7.

[13] Analog Device, Inc., *The Good, the Bad, and the Ugly Aspects of ADC Input Noise—Is No Noise Good Noise ?*, https://www.analog.com/cn/analog-dialogue/articles/adc-input-noise.html.

[14] Analog Device, Inc., Taking the Mystery out of the Infamous Formula, "SNR = 6.02N + 1.76dB," and Why You Should Care, https://www.analog.com/media/en/training-seminars/tutorials/MT-001.pdf.

[15] J. Zhu et al., *Prototype of Integrated Pulse Digitalization and Readout Electronics for CLYC Detector*, in 2018 IEEE Nuclear Science Symposium and Medical Imaging Conference Record (NSS/MIC), Sydney, Australia, Nov 2018, https://doi.org/10.1109/NSSMIC.2018.8824287.